\documentstyle[prl,aps,preprint]{revtex}

\begin{document}

\preprint{ILC/pdec-17oct95/quant-ph/9511003}

\draft


\title{Quantum Error Correction by Coding}

\author{Isaac L. Chuang$^1$ {\em and\/} R. Laflamme$^2$}

\address{\vspace*{1.2ex}
 	\hspace*{0.5ex}{$^1$ERATO Quantum Fluctuation Project} \\
 	Edward L. Ginzton Laboratory, Stanford University,
 		Stanford, CA 94305 \\[1.2ex]
 	$^2$Theoretical Astrophysics, T-6, MS B288	\\
 	Los Alamos National Laboratory, Los Alamos, NM 87545, USA}

\date{\today}
\maketitle



\begin{abstract}
Recent progress in quantum cryptography and quantum computers has
given hope to their imminent practical realization.  An essential
element at the heart of the application of these quantum systems is a
quantum error correction scheme.  We propose a new technique based on
the use of coding in order to detect and correct errors due to
imperfect transmission lines in quantum cryptography or memories in
quantum computers.  We give a particular example of how to detect a
decohered qubit in order to transmit or preserve with high
fidelity the original qubit.
\end{abstract}




Classical information theory tells us that messages may be
communicated with high fidelity and at a finite rate, even along noisy
channels, by using appropriate coding techniques
\cite{Shannon48,Cover91}.  Does this apply for transmitting {\em
quantum} messages along {\em quantum} channels?  Consider the
following scenario: Alice wishes to send some qubit (quantum two-state
system) $c_0\,|{0}\rangle +c_1\,|{1}\rangle $ to Bob, but their
communication link is noisy -- interaction with the line during
transmission causes the information stored in the superposition state
of the qubit to be lost to the environment (Figure~\ref{fig:comlink}).
For example, Alice may prepare an electronic state of a single ion in
some superposition state, then Federal Express it to Bob;
unfortunately, field fluctuations along the way destroy part of the
information stored in the superposition of the $|{0}\rangle $ and
$|{1}\rangle $.  Knowing the specific form of the microscopic
interaction Hamiltonian, can Alice encode a qubit in some way such
that Bob can detect and correct transmission errors?

The answer to this {\em quantum noisy coding problem} is presently
unknown\cite{Schum95,Hausladen95}.  However, we have gained some
insight by studying the special case of
decoherence\cite{Zurek91,Zurek93}: phase damping between the
$|{0}\rangle $ and $|{1}\rangle $ states which leaves the relative
probabilities intact.  We have discovered a coding technique for
partially detecting and correcting errors due to decoherence.  Our
idea is based on using a representation for a quantum bit which is
sensitive to the quantum jumps induced by decoherence.  Furthermore,
the representation is constructed in such a manner that if no jump
occurs, the state is left intact with high probability.  The two cases
are discriminated by using a projective measurement, which selects the
original qubit with high probability, as the amount of redundancy in
the code is increased.  The key to why this works lies in an
understanding of the decoherence process.  The coding will also detect
error due to amplitude damping, the process occurring when the excited
state $|{1}\rangle $ decays to the ground state with a given
probability.


High-fidelity transmission through a noisy channel is based on
knowledge about the noise structure.  For example, in the case of
transmission of classical information where the $|1\rangle$ state can
decay into $|0\rangle$, it is possible to make a code where every bit
of the original message is mapped into a new message using an even
number of $|1\rangle$ states. If we end up with an odd number of
$|1\rangle$ states, we know that we have lost information during
transmission\cite{Cover91}.

The same principle is also true for a quantum channel.  Unwanted
environmental interactions along a channel cause the message
$|{\psi}\rangle $ to {\em decohere} into some mixed state
$\rho_{noisy} = \$ ( |\psi\rangle\langle\psi| )$, where $\$$ is a
superscattering operator which describes the noise
process\cite{Hawking82,Gardiner91}.  It is not easy to find a way to
make a code to correct for this interaction using the density matrix
in the $(|0\rangle,|1\rangle)$ basis.  However, useful hints can be
obtained utilizing an equivalent {\em single wavefunction} picture of
the noise process that is similar to the methods of quantum
trajectories\cite{Carmichael93,Molm93}.  We will denote a phase-damped
quantum trajectory ``wavefunction'' by the subscript ${\it pd}$.
These mixed states will be expressed in the basis $|{\phi_n}\rangle $
in which all initial states remain diagonal during the noise process,
such that $\rho_{noisy} = \sum_n \,|{\phi_n}\rangle \langle{\phi_n}|$.

The main gist of our coding technique is to represent a single qubit
$|{\psi_0}\rangle = c_0 |{0}\rangle + c_1|{1}\rangle $ as some state
$|{\psi}\rangle = c_0 |{0_L}\rangle + c_1 |{1_L}\rangle $ using $N$
qubits.  Decoherence causes us to get a mixture of $|{\phi_n}\rangle $
states.  However, our qubit representation is designed such that
decoherence acts symmetrically upon the whole state, such that with
probability $p_0$ we have $|{\phi_0}\rangle = \sqrt{p_0}
|{\psi}\rangle $, which corresponds to having the qubit remain intact
through the interaction.  For $n\ge 1$, the states $|{\phi_n}\rangle $
describe cases when the qubit symmetry is disrupted by the noise
process.  These are undesirable final states which have to be
rejected.  Thus, if $|{\phi_0}\rangle $ were orthogonal to all the
others, we could detect errors perfectly by distinguishing the two
manifolds.  Although we have not found such a perfect scheme, we can
come close, as we show below.

Let us begin by describing the decoherence process.
An assembly of $N$ qubits, represented by the state
\begin{equation}
	|{\psi}\rangle  = \sum_{b\in{\cal B}} c_b \, |{b}\rangle
\end{equation}
decoheres due to interaction with the environment (with operators
$a^i_{k}$) through the Hamiltonian
\begin{equation}
	H_{I} = \lambda' \sum_{i=1}^N \sum_k \sigma_z^i
			{a^i_{k}}^\dagger a^i_{k}
\end{equation}
into the mixed state described by the density matrix
\begin{equation}
	\rho_{pd} = \$_\Lambda (|{\psi}\rangle \langle{\psi}|)
	= \sum_{a\in{\cal B}} \sum_{b\in{\cal B}}
		c_a c_b^* e^{-\lambda h(a,b)} \, |{a}\rangle \langle{b}|
\,,
\label{eq:rhopd}
\end{equation}
where $\$_\Lambda$ is the superscattering operator, $\lambda$
parameterizes the amount of damping, $\sigma_z^i$ rotates the $i^{th}$
qubit about the computational basis, $a$ and $b$ are binary strings of
$N$ bits, and ${\cal B} = \{0,1\}^N$ is the set of $2^N$ bit-strings
which span the Hilbert space.  $h(a,b) = a\ {\rm xor}\ b$ gives the
Hamming distance\cite{hamming} between $a$ and $b$; it appears because
the rate at which the off-diagonal element $\langle
a|\rho_{pd}|b\rangle$ decays is proportional to the number of bits
different between $a$ and $b$.  This assumes that each qubit is
decohered by an independent bath, which is physically very reasonable.

The mixed state described by the density matrix $\rho_{pd}$ can be
decomposed into the explicit statistical mixture of pure states
\begin{equation}
	\rho_{pd} = \sum_{n=0}^{2^N-1} \,|{\phi_n}\rangle \langle{\phi_n}|
\label{eq:defrpd}
\end{equation}
where
\begin{equation}
	|{\phi_n}\rangle  = \left\{ \begin{array}{l} \displaystyle
	    \sum_{b\in{\cal B}} c_b e^{-\lambda h(b)} \,|{b}\rangle
		\hspace*{16ex} \mbox{for $n=0$}
\\
	    \displaystyle
	    \sum_{b\in{\cal B}}
		\theta(n\wedge b)(1-\theta(n\wedge\overline{b}))\,
\nonumber\\ \hspace*{4ex} \times\,
		c_b \, e^{-\lambda h(n,b)}
		\left[1-e^{-2\lambda}\right]^{\frac{h(n\wedge b)}{2}}
		\,|{b}\rangle
		\hspace*{2ex} \mbox{for $n\ge1$}
	\end{array}\right.
\,.
\label{eq:swpd}
\end{equation}
$\wedge$ denotes the bitwise binary {\sc and} function, $\theta(x)=1$
for $x>0$ and zero otherwise is the usual step function, $h(x) =
h(x,0)$ is the Hamming weight of $x$, and $\overline{b}$ denotes the
bitwise complement of $b$.  The function of $\theta(n\wedge
b)(1-\theta(n\wedge\overline{b}))$ is to select those values of $n$
which have one's in its bit-string only where $b$ does.  Proof of the
equivalence of Eqs.~(\ref{eq:defrpd}-\ref{eq:swpd}) to
Eq.~(\ref{eq:rhopd}) is straightforward, and follows from showing that
$\langle{a}|\rho_{pd} |{b}\rangle = c_a c_b^* e^{-\lambda h(a,b)}$.

In the quantum trajectory picture, the effect of phase damping on the
state $|\psi\rangle$ can thus be described as
\begin{equation}
	|{\psi_{pd}\rangle }
		= \$_\Lambda |{\psi}\rangle
		= \bigoplus_{n=0}^{2^N-1} \,|{\phi_n}\rangle
\,.
\end{equation}
The $\oplus$ denotes a direct sum of the vector spaces (in contrast to
a tensor product), such that $|\langle\alpha|\psi_{pd}\rangle|^2 =
\sum_n |\langle\alpha|\phi_n\rangle|^2$ for an arbitrary pure state
$\langle\alpha|$.  For $p_n = |\langle\phi_n|\phi_n\rangle|^2$, we
have that $\sum_n p_n = 1$, so we may understand $p_n$ as being the
weight of $|{\phi_n}\rangle $ in the mixture, and $|{\phi_n}\rangle
/p_n$ as the possible final pure states after the decoherence.
Physically, one may think of the decoherence process as occurring
because of phase randomization due to interaction with a bath
coordinate.  In this picture, the $n=0$ state results when the
interaction leaves the coordinate unchanged; otherwise, one of the
$n>1$ states results.  Keeping with the quantum trajectories idea, in
the former case it can be said that the wavefunction is rotated by a
non-unitary transform, and in the latter case, a quantum jump occurs.

The single wavefunction approach helps us in devising a code because
we can think of the different results of the transmission as vectors
in the Hilbert space, instead of having to use the space of density
matrices.  Geometrically, our coding technique works by first
extending the Hilbert space by including ancilla qubits.  A single
qubit is then coded with the help of the ancilla in such a way that
when the system has decohered we can decode and project the state so
that the final qubit is as near to the original as possible.

The standard representation of a qubit encodes the logical zero and
one states as $|{0_L}\rangle  = |{0}\rangle $ and $|{1_L}\rangle  =
|{1}\rangle $, such
that an arbitrary qubit is given by the state
\begin{equation}
	|{\psi}\rangle  = c_0 \,|{0}\rangle  + c_1 \,|{1}\rangle
\,.
\label{eq:normrep}
\end{equation}
Using Eq.~(\ref{eq:swpd}), we find that phase damping turns this pure
state into the mixture
\begin{equation}
	|{\psi_{pd}}\rangle
	=
		c_0 \,|{0}\rangle  + c_1 e^{-\lambda} \,|{1}\rangle
	\ \oplus\
		c_1 \left[ 1-e^{-2\lambda} \right]^{1/2} \,|{1}\rangle
\,.
\end{equation}
Using the nomenclature of \cite{Molm93}, decoherence either leads the
original state to be rotated non-unitarily, or to a quantum jump into
$|{\phi_1}\rangle $ (i.e. into the $|{1}\rangle $ state).  The latter
results with probability $|c_1|^2 (1-e^{-2\lambda})$.  Unfortunately,
(1) the $|{1}\rangle $ state is in the space spanned by
$|{\psi}\rangle $ and thus there is no way of detecting a jump, and
(2) even if no jump occurs, the damping has deformed the original
state and so $|{\psi}\rangle $ cannot be recovered intact.

Consider instead a single qubit which is represented by a sequence of
$N$ qubits, with the help of $N-1$ ancilla.  Specifically, let ${\cal
C} = \{0\cdots001, 0\cdots010, \ldots, 10\cdots0\}$ be the set of all
length $N/2$ bit-strings containing only one $1$, such that we have
the representation
\begin{eqnarray}
	|{0_L}\rangle  &=& \sqrt{\frac{2}{N}} \sum_{b\in{\cal C}}
|{0\cdots0b}\rangle
\label{eq:codeA}
\\	|{1_L}\rangle  &=& \sqrt{\frac{2}{N}} \sum_{b\in{\cal C}}
|{b0\cdots0}\rangle
\,,
\label{eq:codeB}
\end{eqnarray}
where the label of each ket contains $N$ digits.  We shall call the
manifold defined by these two unit vectors the {\em representation
manifold}, and say that as long as a state lives within this plane, it
satisfies the {\em representation invariance} condition.  The effect
of phase damping on an arbitrary qubit superposition, $|{\psi}\rangle  =
c_0\, |{0_L}\rangle  + c_1\,|{1_L}\rangle $ is found to be
\begin{eqnarray}
	|{\psi_{pd}\rangle } &=& e^{-\lambda} \,|{\psi}\rangle  \ \oplus\
		\left[ \bigoplus_{b\in{\cal C}} c_0 \sqrt{\frac{2}{N}}
			\left[{1-e^{-2\lambda}}\right]^{1/2}
			\,|{0\cdots0b}\rangle \right]
\nonumber\\ &&\hspace*{4ex}
		\ \oplus\
		\left[ \bigoplus_{b\in{\cal C}} c_1 \sqrt{\frac{2}{N}}
			\left[{1-e^{-2\lambda}}\right]^{1/2}
			\,|{b0\cdots0}\rangle \right]
\,.
\label{eq:psipd}
\end{eqnarray}
This time, when a jump occurs, it results in some state
$|{0\cdots0b}\rangle $ or $|{b0\cdots0}\rangle $ for some $b\in{\cal
B}$, and furthermore, this result contains a component orthogonal to
the representation manifold.  It thus violates the representation
invariance condition, meaning that we can detect, with probability $1
- |\langle0_L|\psi_{pd}\rangle|^2 - |\langle1_L|\psi_{pd}\rangle|^2$
when any jump has occured, using a projective measurement (which
leaves the qubit intact).  Furthermore, when a jump does not occur,
because of the symmetry of the effect of decoherence on states in the
representation manifold, the original qubit is left intact!  This
result is the basis for an error correction scheme against
decoherence.

Note that in our model, we do not assume that the ancilla are
``error-free.''  Rather, the ancilla qubits decohere along with the
original qubit; this is important because that is the case for
realistic systems.


The qubit code given in Eqs.~(\ref{eq:codeA}-\ref{eq:codeB}) can be
used to increase the probability of successful transmission of a qubit
through an imperfect communication link (Figure~\ref{fig:pdecscheme}).
Alice prepares her single qubit $|{\psi_0}\rangle = c_0 \,|{0}\rangle
+ c_1 \,|{1}\rangle $, and introduces $N-1$ ancilla qubits (prepared
in the ground state $|{0\cdots0}\rangle $) to get $|{\psi_1}\rangle =
c_0 \,|{e_0}\rangle + c_1 \,|{e_1}\rangle $, where $|{e_0}\rangle =
|{0\cdots00}\rangle $ and $|{e_1}\rangle = |{0\cdots01}\rangle $.
This is fed into the unitary coding transform $U$ to give
$|{\psi_2}\rangle = U |{\psi_1}\rangle = c_0 \,|{0_L}\rangle + c_1
\,|{1_L}\rangle $.  For example, we may have
\begin{eqnarray}
	|{\psi_2}\rangle  &&= \frac{c_0}{\sqrt{3}}
	     \left[|000001\rangle + |000010\rangle + |000100\rangle \right]
\nonumber\\ &&\hspace*{4ex}
		+ \frac{c_1}{\sqrt{3}}
	     \left[|001000\rangle + |010000\rangle + |100000\rangle \right]
\,.
\end{eqnarray}
The result is transmitted to Bob, who decodes his received mixed state
$|{\psi_3^{pd}}\rangle = \$_\Lambda |{\psi_2}\rangle $ to get
$|{\psi_4^{pd}}\rangle = U^\dagger \,|{\psi_3^{pd}}\rangle $.  Note that
$|{\psi_3^{pd}}\rangle $ is given by Eq.(\ref{eq:psipd}).  Bob will be
interested in two probabilities.  He will reject the entire
transmission if any ancilla qubit is measured to be nonzero;
otherwise, he will accept it.  This happens with probability
\begin{eqnarray}
	P_{\rm accept} &=&
	|\langle e_0|\psi_4^{pd}\rangle|^2 + |\langle e_1|\psi_4^{pd}\rangle|^2
%
%
\\
      &=& \frac{2}{N} + \left(1-\frac{2}{N}\right) e^{-2\lambda}
\,.
\label{eq:paccept}
\end{eqnarray}
As $\lambda\rightarrow \infty$, $P_{\rm accept}\rightarrow 2/N$
because even when the state becomes completely decohered, there is
some probability of not detecting the error.  Note that for small
$\lambda$, the rejection rate $(1-P_{\rm accept})^{-1}$ is essentially
independent of $N$.  When all the $N-1$ ancilla bits are found to be
zero, then the qubit Bob receives is a ``good'' qubit, which is
described by the density matrix
\begin{eqnarray}
	\rho_5 &=& \frac{1}{P_{\rm accept}}
	\left[	\begin{array}{cc}
		|\langle0_L|\psi_3\rangle|^2
			& \langle0_L|\psi_3\rangle\langle\psi_3|1_L\rangle \\
		\langle1_L|\psi_3\rangle\langle\psi_3|0_L\rangle
			& |\langle1_L|\psi_3\rangle|^2
	\end{array}	\right]
\\
	&=&
	\left[	\begin{array}{cc}
		|c_0|^2		& J c_0 c_1^*	\\
		J c_0^* c_1	& |c_1|^2
	\end{array}	\right]
\,,
\end{eqnarray}
where $J$ describes the decoherence which occurs despite the error
correction scheme, and is found to be
\begin{equation}
	J = \frac{N}{2e^{2\lambda}-2+N} \approx 1-\frac{4\lambda}{N}
\,.
\label{eq:J}
\end{equation}
For small $\lambda$, the amount of decoherence suffered decreases
inversely as the number of qubits $N$ used in the code.  In
comparison, if the usual qubit representation Eq.~(\ref{eq:normrep})
is used, the amount of decoherence suffered is $J_0 = e^{-\lambda}
\approx 1-\lambda$, and thus the advantage of our scheme is that it
causes the off-diagonal terms to decay less quickly, as long as
$N>2(1+e^\lambda)$ (recall that $e^{\lambda}$ is the decoherence
suffered by only one qubit).  Finally, Bob extracts the correct result
with probability
\begin{equation}
	P_{\rm correct} = |\langle \psi_0| \rho_5 |\psi_0 \rangle|^2
	= 1-2|c_0|^2|c_1|^2 (1-J)
	\geq \frac{1+J}{2}
\,,
\end{equation}
so that $P_{\rm correct} \geq 1-2\lambda/N$ for small $\lambda$.  This
probability is known as the transmission fidelity ${\cal F} \cite{Schum95}$.

One interesting question to ask is: in analogy to the watchdog
effect\cite{Zurek84}, can the error detection probability and
transmission fidelity be improved by periodic correction?  The answer
depends on the form of the errors suffered as a function of time.
Instead of suffering decoherence $\lambda$, we may perform
$k$ corrections each with decoherence $\lambda/k$.  In this case, we
find that
\begin{equation}
	P_{\rm accept} = \left[
		\frac{2}{N}+\left(1-\frac{2}{N}\right) e^{-\lambda} \right]^k
\hspace*{1ex} {\rm and} \hspace*{1ex}
	J_k = \left[ \frac{N}{2e^{\lambda}-2+N} \right]^k
\,,
\end{equation}
which, unfortunately, is worse than the result of
Eqs.(\ref{eq:paccept}) and (\ref{eq:J}) for any $N>2$.  Of course,
this happens because we have assumed that phase damping occurs
exponentially with time, in which case it is known that the watchdog
effect is ineffective.  Instead, of $e^{-\lambda t}$, if we have an
error rate which is quadratic in time, $1-\epsilon t^2$, then we must
compare
\begin{equation}
	J = \frac{N}{2(1-k^2\epsilon)-2+N}
\hspace*{2ex}{\rm and}\hspace*{2ex}
	J_k = \left[ \frac{N}{2(1-\epsilon)-2+N} \right]^k
\,.
\end{equation}
For small error per step $\epsilon$, $J\approx
1+2k^2\epsilon/N$, and $J_k \approx 1+2k\epsilon/N$, so we find that
periodic correction {\em is} effective.  Our scheme works hand-in-hand
with the principle of watchdog stabilization.


By increasing the size of the Hilbert space and using a coding which
distributes unwanted transmission errors symmetrically, we have
demonstrated how a single qubit can be coded to guarantee as high
transmission fidelity as desired using additional ancilla bits.
Another interesting characteristic about the scheme presented here is
that it can also be used to perfectly detect errors due to amplitude
decoherence.  The coding in Eqs.~(\ref{eq:codeA}-\ref{eq:codeB}) is a
generalization of the dual-rail bit of \cite{Chuang94,Chuang95b} and
their scheme can be adapted straightforwardly.  Our result may be
contrasted with that of \cite{Shor95b}.  Shor has a scheme that is
able to reconstruct the initial state exactly assuming that only one
of nine bits decohered. Our scheme is independent of the decoherence
strength.  It might be possible to adapt Shor's scheme to ours in
order to get perfect fidelity.  Bennett et. al. have devised a method
to rejuvenate EPR pairs that have lost their purity\cite{bbssw95}.
Their method uses EPR bits to accomplish this process, while ours
needs only ancilla qubits in their ground states and will be
advantageous when EPR pairs are expensive.

We believe that other (better) coding schemes against decoherence,
developed along the lines we have presented, may exist.  In our
search, we found an intriguing one using only one ancilla where
\begin{eqnarray}
	|{0_L}\rangle  &=& \frac{ |{00}\rangle  +|{11}\rangle  }{\sqrt{2}}
\\	|{1_L}\rangle  &=& \frac{ |{01}\rangle  +|{10}\rangle  }{\sqrt{2}}
\,,
\label{eq:codetwo}
\end{eqnarray}
for which the probability of acceptance and the fidelity are
\begin{equation}
	P_{\rm accept} =  \frac{1+e^{-2\lambda}}{2}
\hspace*{4ex} {\rm and} \hspace*{4ex}
	{\cal F} = 1-2 \frac{|c_0|^2|c_1|^2}{\cosh\lambda}
\,.
\end{equation}
The interesting point here is that for small $\lambda$, ${\cal F}$ is
quadratic in $\lambda$.  However we were unable to generalize this to
a scheme which would go as $\lambda^2/N$ for a $N$-bit code.  This is
in contrast with the code in Eqs.~(\ref{eq:codeA}-\ref{eq:codeB}).

In conclusion, our error correction technique uses a $N$-qubit
representation of a single qubit to increase the transmission fidelity
through a noisy quantum channel from $1-\lambda/2$ (for no error
correction) to at least $1-2\lambda/N$ for the accepted qubit,
using the code in
Eqs.~(\ref{eq:codeA}-\ref{eq:codeB}).  This result provides an example
of how coding can be used to construct representations which are
robust against phase decoherence.  The same general technique may be
applied to construct error correcting codes for other sources of
decoherence.




\onecolumn

\begin{figure}[p]
\caption{Transmission of a single qubit through a noisy quantum channel.}
\label{fig:comlink}
\end{figure}

\begin{figure}[p]
\caption{Single qubit transmission using a code which is robust
	 against phase decoherence.  The $N=4$ case is pictured.}
\label{fig:pdecscheme}
\end{figure}

\end{document}